# Tumor Microenvironment-based Gene Signatures Divides Novel Immune and Stromal Subgroup Classification of Lung Adenocarcinoma

Zihang Zeng[1,2,3], Jiali Li[1,2,3], Nannan Zhang[1,2,3], Xueping Jiang[1,2,3], Yanping Gao[1,2,3], Liexi Xu[1,2,3], Xingyu Liu[1,2,3], Jiarui Chen[1,2,3], Yuke Gao[1,2,3], Linzhi Han[1,2,3], Jiangbo Ren[4], Yan Gong[4,*], Conghua Xie[1,2,3,*]

[1] Department of Radiation and Medical Oncology, Zhongnan Hospital of Wuhan University, Wuhan, Hubei, 430071, The People's Republic of China

[2] Hubei Key Laboratory of Tumor Biological Behaviors, Zhongnan Hospital of Wuhan University, Wuhan, Hubei, 430071, The People's Republic of China

[3] Hubei Cancer Clinical Study Center, Zhongnan Hospital of Wuhan University, Wuhan, Hubei, 430071, The People's Republic of China

[4] Department of Biological Repositories, Zhongnan Hospital of Wuhan University, Wuhan, Hubei, 430071, The People's Republic of China

**Running title:** TME Genes Divides Novel Lung Adenocarcinoma Classification

**Additional information:** This study was supported by National Natural Science Foundation of China [81372498, 81572967, 81773236, and 81800429]; National Project for Improving the Ability of Diagnosis and Treatment of Difficult Diseases, National Key Clinical Speciality Construction Program of China [[2013]544]; the Fundamental Research Funds for the Central Universities [2042018kf0065 and 2042018kf1037]; Health Commission of Hubei Province Scientific Research Project [WJ2019H002 and WJ2019Q047]; Wuhan City Huanghe Talents Plan, and Zhongnan Hospital of Wuhan University Science, Technology and Innovation Seed Fund [znpy2016050, znpy2017001, znpy2017049, and znpy2018028].

*Correspondence author: Conghua Xie. Tel: +86-27-6781-2607; Fax: +86-27-6781-2892; Email: chxie_65@whu.edu.cn. Correspondence may also be addressed to Yan Gong. Tel: +86-27-67811461 ; Fax: +86-27-67811471; Email: yan.gong@whu.edu.cn

There is no conflict of interest in all authors.

## Abstract

Tumor microenvironment has complex effects on tumorigenesis and metastasis. However, there is still a lack of comprehensive understanding of the relationship among molecular and cellular characteristics in tumor microenvironment, clinical prognosis and immunotherpy response. In this study, the immune and stromal (non-immune) signatures of tumor microenvironment were integrated to identify novel subgroups of lung adenocarcinoma by eigendecomposition and extraction algorithms of bioinformatics and machine learning, such as non-negative matrix factorization and multitask learning. Tumors were classified into 4 groups according to the activation of immunity and stroma by novel signatures. The 4 groups had different mutation landscape, molecular, cellular characteristics and prognosis, which have been validation in 6 independent data sets containing 1551 patients. High-immune and low-stromal activation group links to high immunocyte infiltration, high immunocompetence, low fibroblasts, endothelial cells, collagen, laminin, tumor mutation burden, and better overall survival. We developed a novel model based on tumor microenvironment by integrating immune and stromal activation, namely PMBT (prognostic model based on tumor microenvironment). The PMBT showed the value to predict overall survival and immunotherapy responses. We submitted an simple R package (https://github.com/ZengZihang/PMBT) to enable any researcher to apply our novel immune, stromal and PMBT scores.

## Introduction

Lung cancer is the most frequent tumor, of which non-small cell lung cancer (NSCLC) accounts for 80 percent (1, 2). Lung adenocarcinoma (LUAD) is well acknowledged for its malignancy and morbidity rate among people, and exhibits diversity of gene mutations. Normally, innate and required immunity are responsible for tumor cell distinguishment and elimination in tumor immune microenvironment (3). With the discovery of immune checkpoint inhibitors, such as programmed death 1 (PD-1) and its ligand PD-L1 (PD-1/PD-L1) (4), cytotoxic T-lymphocyte-associated protein 4 (CLTA-4), immunotherapy becomes a promising method to treat LUAD (5). Nevertheless, immunotherapy only benefit ~16% LUAD patients for long-term survival (6). It remains unsolved to deal with patients with therapeutic tolerance (7). Tumor stroma (non-immune) was reported to be closely related to the progression, metastasis and poor prognosis of tumors (8). However, the immune and stroma components of tumor microenvironment (TME) are to be investigated. This study aimed to propose a new insight and classification method to integrate both immune and stromal impacts on TME in LUAD.

TME is a very complex mixture, containing tumor cells, endothelial cells of blood and lymphatic vessels, fibroblasts, immune cells and normal tissues (9). In the common analysis such as gene chip and second generation sequencing, the results may be unstable due to the influence of the TME, since the non-tumor characteristic signals are not identified. Therefore, this study tried to identify the characteristic genes of TME by signal decomposition algorithm. Non-negative matrix factorization (NMF) is a practical approach to decomposite matrix and distract abstract characteristics among massive data sets, which is commonly exploited for gene pattern recognition and computer vision in biomedical engineering (10, 11). Multitask learning (MTL) is an inductive transfer method to improve predictive capability by learning tasks in parallel while using shared representations (12, 13).

Integrating multi-omics data helps to achieve a comprehensive understanding in medical research (14). Besides transcriptionomics analysis, genomics data were used to calculate the tumor mutation burden (TMB) and frequency of single gene mutation in this study.

In general, we identified the 4 subgroups based on TME features in LUAD. Different molecular, cellular characteristics, mutation landscape and prognosis were discovered in the 4 groups, which was validated in the 6 independent data sets containing 1551 patients. The flow chart of this study was showed in Supplementary Fig. S1.

## Materials and Methods

### Data standardization and preprocessing

Six data sets containing transcriptome expression profile and clinical information of 1551 LUAD patients containing 1 train data set from TCGA-LUAD and the 5 validation data sets from GEO database (Supplementary Table S1). Oligoexpression genes not expressed in most samples were excluded from analysis. The high throughput transcriptome sequencing data (TCGA-LUAD) implemented RNA-Seq by Expectation Maximization (RSEM) and log2(x+1) standardization (15), and RNA profiling chip (GSE11969, GSE68465, GSE68571, GSE37745 and GSE50081) adopted robust multi-array average (RMA) standardization (16-22). All data were normalization by Z-score and min-max methods to further analysis.

### Non-Negative Matrix Factorization

Unsupervised low-dimensional microdissected feature was extracted from TME by NMF method (23). The K number of NMF factors (K = 11) was determined by hierarchical clustering of TCGA transcriptome data (Supplementary Fig. S2A). Primitive gene expression matrix was decomposed into gene loading matrix and sample loading matrix through brunet methods after

2000 iterations (24). Top 100 weighted genes of each NMF factor were considered as characteristic genes for subsequent enrichment analysis.

**Calculating the Levels of Different Immune and Stromal Features in TME**

TME related cell content was examined by MCPcounter (8 types of immune cells, endothelial cells and fibroblasts, https://zenodo.org/record/61372), CIBERSORT (22 types of human hematopoietic cells, http://cibersort.stanford.edu) and TIMER (6 types of immune cells, https://cistrome.shinyapps.io/timer/) in each sample (25-27). The Pearson correlation analysis of cell contents in different software was used to find robust results.

The single-sample gene set enrichment analysis (ssGSEA) for microarray and RNA-Seq data was used to calculate the enrichment score of each sample based on different immunity related and stroma related genes in TME (28). The production activity of interleukin-1,2,6,10 and IFN-γ was quantitatively calculated by ssGSEA using gene features from Gene Ontology (GO, http://www.geneontology.org) (29). PD-1 signaling activation was calculated by Quigley's PD-1 related genes (30). The tumor-infiltrating lymphocytes (TIL) scores of hematoxylin and eosin (H&E) sections from TCGA were obtained from previous study (31). The immune and stromal scores were calculated by ssGSEA using Kosuke's immune and stromal characteristic genes (32) and Richard stromal characteristic genes (9).

**ℓ2,1-Norm Regularized Multitask Learning**

MTL was used to filter genes related to multiple traits. This method was beneficial for making the results more sparse and robust by limit of ℓ2,1-norm regularization. MTL consists of multiple linear regression sub-tasks containing NMF factor feature and TME cell relative contents. By optimizing the ℓ1 and ℓ2 norm, a small number of genes with regression coefficients in all subtasks were screened out. In this study, 5 subtasks were involved in MTL including 2 immune and stromal NMF factor features, T cell, B cell, endothelial cell and fibroblast abundance of MCPcounter. The code of MTL was based on MATLAB (https://github.com/jiayuzhou/MALSAR).

**GO Enrichment and Gene Set Enrichment Analysis**

The GO enrichment included over-representation analysis (ORA) and gene set enrichment analysis (GSEA) using WebGestalt tools (http://www.webgestalt.org/option.php) and R clusterProfiler package (33, 34).

**Different Gene Expression Analysis**

Different gene expression (DGE) analysis was used to identify different expression genes (DEGs) with significant difference at RNA level between the 2 classified phenotype, which was realized by R limma package (35).

**Mutation Analysis**

LUAD mutation data were obtained from whole-exome sequencing files on Genomic Data Commons website (https://portal.gdc.cancer.gov/repository). Mutation data were analyzed using maftools package of R language (36).

**Meta-Analysis**

For meta-analysis, our prognostic effect of interest was overall survival (OS). Hazard ratio (HR) integration of TME phenotypic groups with multiple data sets was accomplished using meta package of R (https://cran.r-project.org/doc/Rnews/Rnews_2007-3.pdf). The fixed effect model was used when there was no obvious heterogeneity among multiple data sets ($I^2 < 50\%$), otherwise the random effect model was used. The HR and 95% confidence interval (CI) were visualized in forest plot. The 3,5-year survival rates were calculated using a non-controlled method. The standard error of survival rates were estimated as binary variables. The calculation was as follows:

$$SE(Pt) = \sqrt{Pt^\wedge 2 * (1 - Pt)/Ns}$$

where Pt is survival rate of time t, SE(Pt) is the standard error of Pt, and Ns is the number of risk at time t. Non-controlled meta analysis was realized by Importing SE(Pt) and Pt into Revman 5 software (https://community.cochrane.org/help/tools-and-software/revman-5).

**Intercellular Communication Network**

The intercellular communication networks were constructed by priori information of cell-receptor, cell-ligand and receptors-ligand pairs from the FANTOM5 resource (http://fantom.gsc.riken.jp/5/suppl/Ramilowski_et_al_2015/) (37). The TME cell types in MCPcounter did not completely coincide with those in FANTOM5. The abundance of endothelial cells and fibroblasts was calculated by MCPcounter algorithm for all types of endothelial cells and fibroblasts in FANTOM5. Receptor ligand characteristics were defined as intersection characteristics of endothelial cells and fibroblasts of FANTOM5, after removing outlier cell types in FANTOM5 with low abundance by MCPcounter. Similar to Thorsson's study (31), the edges in each TME subtype were determined in the scaffold of possible interactions. Cell population, ligand and receptor gene expression values were divided into 3 parts (low, medium and high). If at least 66% of the samples in the subtype were mapped to medium or high value of one node, this node was input into the subtype network. Only when the 2 nodes consisting in the edge were in the network, can the edge and node be incorporated into the network. Core subnetworks in different subtype networks were determined by Molecular Complex Detection (MOCDE) method of Cytoscape (38).

**Statistical analysis**

All statistical analysis was implemented by R 3.5.1 (https://www.r-project.org/). R survival package was used to cox proportional hazards regression. Receiver operating characteristic curve (ROC) drawing and (area under curve) AUC calculation were based on pROC packages (39). R stats package was used to Pearson correlation, variance analysis and chi square test. P value less than 0.05 was considered statistically significant in all hypothesis tests. All the P value were two-sided. The significance threshold of false discovery rate (FDR) q value was also 0.05 in GO and KEGG enrichment analysis.

## Results

**The high-immune and low-stromal group and the low-immune and high-stromal group were identified based on immune and stromal activity in TME**

The Pearson correlation analysis of MCPcounter, CIBERSORT and TIMER showed the greater difference between CIBERSORT and TIMER, and MCPcounter showed a higher correlation with the other 2 tools (Supplementary Table S2). Therefore, data from MCPcounter were used for subsequent analysis.

In NMF analysis, we confirmed 4 TME-related NMF factors with different immune and stromal enrichment score as reported previously (Supplementary Fig. 2B) (9, 32). Interesting, for univariate Cox regression, the significant positive relationship between NMF factor expression feature and prognosis was in high-immune and low-stromal enrichment score group (n = 35, P = 0.000144). Therefore, there were no significant effect in both double high (n = 24, P = 0.721) and double low group (n = 31, P = 0.292). The low-immune and high-stromal group showed adverse effects on OS (n = 12, P = 0.1524). The survival curves of high-immune and low-stromal group versus the low-immune and high-stromal group were also different (Figure 2C, HR = 0.5597458, P = 0.4776). The high-immune and low-stromal activity expression pattern (NMF factor 6, Supplementary Fig. 2B) was identified. The enrichment analysis of the exemplar genes (Top 100 weighted) in NMF factor 6 shows high immune activity including humoral immune response (P = 1.37853438476176E-11, q-value = 1.05E-11), T cell activation (P = 4.55E-10, q-value = 3.47E-10), response to IFN-γ (P = 4.82E-09, q-value = 3.67E-09) and antigen processing and presentation of exogenous peptide antigen via MHC class II (P = 9.02E-09, q-value = 6.87E-09, Supplementary Table S3, Supplementary Fig. 2D&E). In addition, significant stromal activation contained extracellular matrix organization (P = 9.84E-62, q-value = 7.30E-62), collagen fibril organization (P = 9.84E-62, q-value = 7.30E-62), response to transforming growth factor beta (P = 9.84E-62, q-value = 7.30E-62, Table S4, Supplementary Fig. 2F&G) for low-immune and high-stromal activity expression pattern (NMF factor 5, Supplementary Fig. 2B).

Because the above TME-related NMF factors were low-dimensional and coarse-grained, to further identify genes related to immune and stroma, we used MTL algorithm to screen genes associated with 6 subtasks (feature of NMF factor 5, feature of NMF factor 6, T cell, B cell, endothelial cell and fibroblast abundance of MCPcounter). The 179 immune and stroma related genes were identified including many cluster of differentiation molecules and ligands (CD1A, CD1B, CD1E, CD3G, CD8A, CD19, CD74, CD84, CD163, CD40LG, CD300LG), IFNG (IFN-γ), VEGFA (vascular endothelial growth factor A), ITGB2 (integrin beta-2, CD18), major histocompatibility complex molecule (HLA-DQA1, HLA-DPB1, HLA-DPA1, HLA-DRB1, HLA-DRA, HLA-DQA2, HLA-B, HLA-C and HLA-A ), chemoattractants (CCL5, CCL19) and collagen related genes (COL1A1, COL1A2, COL3A1, COL6A6, COL11A1 and COL29A1). Consensus clustering (K = 2) of the above 179 genes successfully divided the patients into the high immune-stromal ratio group and the low immune-stromal ratio group (Supplementary Figure S3). DEGs were further analyzed between the above 2 groups (Supplementary Fig. S4, Supplementary Table S5) for GSEA. Predictably, the high immune-stromal ratio group had varied interleukin levels, IFN-γ production and adaptive immune activation, and the low immune-stromal ratio group had higher oxidative phosphorylation, double-strand break repair activation (P < 0.001, Supplementary Fig. S5, Supplementary Table S6).

**Novel TME gene signatures and enrichment score**

Novel TME gene signatures were optimized using DGE analysis for the high immune-stromal ratio group versus the low immune-stromal ratio group. DGE analysis selected 166 DEGs with P value less than 0.05 from the 179 immune and stromal characteristic genes (Supplementary Table S7), and divide them into 2 categories (108 immunity-related category genes and 58 stroma-related category genes, Fig. 1A). Novel immune and stromal scores were calculated by immunity-related and stroma-related genes. The new scores revealed higher relevance with TME cells (Supplementary Fig. S6) and closer links to OS (immune score, beta = -1.7008, P = 0.0038; stromal score, beta = 7.245, P = 3.29e-05) than previous scores (Kosuke's immune score, beta = -0.8936, P = 0.0276; Kosuke's stroma score, beta = -0.4914, P = 0.2833; Richard's stroma score, beta = 0.6588, P = 0.0859) in Cox regression.

**Cellular and molecular characteristics in the 4 TME groups based on immune and stroma related gene features**

Different immune and stromal activation divided the patients into the 4 categories (HL: high-immune and low-stromal; LH: low-immune and high-stromal; HH: double high; LL: double low). The 4 classifications were successfully achieved through the median value of the immune and stromal scores. These 4 subtypes showed distinct TME-related gene expression patterns (Fig. 1B).

Extensive hyperimmune cell infiltration and TIL were found in the hyperimmune class (P < 0.01, Fig. 2C). However, the high stromal activity group showed a large number of fibroblast infiltration (P < 2.2e-16, Fig. 1C). To further understand the biological characteristics of the 4 subtypes at the whole mRNA levels, DGE analysis was used to identify high-expression gene signatures of different classes (Fold change > 0, adjusted P-value < 0.05). In the HH group, T cell activation (P = 0, q-value = 0.0010290), humoral immune response (P = 0, q-value = 0.00045020) and extracellular structure organization (P = 0, q-value = 0.0044475) were exhibited in GSEA (Fig. 1D, Supplementary Table S8). There were significant enrichment in the cytoskeleton, cell cycle regulation and DNA repair including microtubule cytoskeleton organization involved in mitosis (P = 0, q-value = 5.33E-04) and double-strand break repair (P = 0, q-value = 0.04247535) in the LH group (Fig. 1D, Supplementary Table S9). Significant immune activation containing B cell activation (P = 0, q-value = 0), T cell activation (P = 0, q-value = 0), interferon-gamma production (P = 0, q-value = 0.00517077) and interleukin-12 production (P = 0, q-value = 0.006059496, Fig. 1D, Supplementary Table S10) was observed in the HL group. No significant enrichment was found in the LL group (q-value > 0.05). The molecular characteristics of different immune and stromal class were shown in Fig. 1E. Patients with high immune score were related to more activation in immune molecules, such as IFN-γ, IL-1, IL-2, IL-6, IL-10 (P < 2.2e-16). Besides, there was higher PD-1 activation in patients with higher immune enrichment scores (P < 2.2e-16).

Six identified pan-cancer immune subtypes reported previously were integrated into our novel classification (31). We found that 89.7% LUAD patients were identified as wound healing (18.3%), IFN-γ dominant (31.7%) and inflammatory (39.7%) subtypes (Supplementary Table S11). Significant higher proportion of wound healing subtypes was shown in the LH group compared others (45.2% vs. 5.3%, P < 2.2e-16), suggesting high proliferation rate, high expression of angiogenic genes and Th2 cell bias. HH group had high proportion of IFN-γ dominant subtypes (59.7% vs. 26.3%, P = 5.311e-08) linked with high M1/M2 macrophage polarization, CD8 signal (Fig. 1E). Inflammatory class had a larger number than other groups in both HL and LL groups (66.0% vs. 54.2% vs. 16.5%, P < 2.2e-16), suggesting low to moderate tumor cell proliferation, high Th17 and Th1 genes. Predictably, the 2 low immune groups had more lymphocyte depleted subtype patients (7.0% vs. 1.4%, P = 0.006637) with Th1 suppression and high M2 response.

**Mutation Landscape and tumor mutation burden of different immune and stromal class**

Mutation analysis was performed to gain the comprehensive perspective of genomic characteristics though different immune and stromal class. Based on bi-classification method (NMF consistency clustering for the above 166 TME characteristic genes), the high immune-stromal ratio group showed a higher number of mutations (median patient mutations 225 vs. 115, P < 0.0001), higher tumor mutation burden (difference between means -3.049 ± 0.6019, P < 0.0001) and a higher mutation rate and Variant Allele Frequency (VAF) in the driving genes (Supplementary Fig. S7 & 8, Fig. 2A-C). For instance, the low immune-stromal ratio patients had high TP53 mutations (55% vs. 38%, P = 0.0002036) and more frame shift deletion (13.46% vs. 8.60%, P = 0.341) and less nonsense mutation (16.03% vs. 22.58%, P =

0.2625). Similarly, higher mutation rate and VAF of KRAS (29% vs. 26%, 81 vs. 60, P = 0.518) were observed in the same groups, and almost all mutations were missense. In the volcanic map (Supplementary Table S12, Fig. 2D), we identified genes related to higher mutation frequency and higher mutation rate difference between the high and low immune-stromal ratio groups. In addition to TP53, COL11A1 and KEAP1 also showed a large rate difference (COL11A1: rate difference = 14.55%, mutation frequency = 19.4%; KEAP1: rate difference = 10.82%, mutation frequency = 17.8%), and they had biological effects of extracellular matrix organization and MHC-mediated antigen processing and presentation in Reactome pathways database (40). The co-occurrence and mutually exclusive mutations were shown in Fig. 2E & F. Moreover, higher mutation of WNT related genes was observed in the low immune-stromal ratio groups (Supplementary Fig. S9 & 10).

For the 4 subgroups, significant lower TMB were found in the HH and LH groups (8.26243 vs. 4.98783, P = 5.637e-08, Fig. 2G). However, there was no significant difference between the HH versus LH and HL versus LL groups (P of pairwise comparisons using t tests: 0.90422 and 0.79067). Actually, high stromal score showed the positive correlation with TMB (cor = 0.2935904, P= 1.619e-11, Fig. 2H). Moreover, there was a strong correlation between the neoantigen of single nucleotide variation and TMB (cor = 0.9549101, P < 2.2e-16, Fig. 2I). Weak negative correlation between immune score and TMB was found in this study (our immune score: cor = -0.1369304, P= 0.002021; previous immune score: cor = -0.1068824, P= 0.01616).

**The prognosis of different subtypes was significantly different**

The clinical characteristics of the 4 groups were shown in the Table 1. Higher pathological Tumor (T) stage was found in the low immune groups (P < 0.01), similar to pathological Node (N) stage (P < 0.05) and pathological tumor stage (P < 0.01), suggesting a link between Tumor Node Metastasis (TNM) stages and immune cell activity in TME.

Different TME subgroups showed significant different prognosis. Clinicopathological information was used to obtain a more comprehensive perspective of prognostic factors for both univariate and multivariate Cox regression analysis. Pathological T stage (β = 0.4153, P = 0.0115) and tumor status (β = 1.0658, P = 1.48e-05) were retained as significant prognostic factors in multivariate Cox regression for backward selection. Both of TNM stages and residual tumors were risk factors for OS (Table S13).

According to Cox regression analysis of TME relative cell contents in TCGA-LUAD cohort, the positive effect of immune cells (T cells, B cells and myeloid dendritic cells) on prognosis and the negative effect of stromal cells (fibroblasts) on prognosis were clarified (Supplementary Table S14). The 2 grouping methods were implemented via consistency clustering of 166 immune and stromal characteristic gene NMF algorithm (K = 2). The dichotomy of LUAD (the high an low immune-stromal ratio groups) showed significant prognostic differences of OS (HR = 0.5597458, P = 2e-04, Fig. 3A). For the 4 subgroups, HL patients had significantly better OS than the others (HR = 0.4617451, P = 3.015633e-05, Fig. 3B). LH patients had worse OS (HR = 1.788947, P = 8.381535e-05), and HH and LL patients were comparable (HR = 1.045836, P = 0.8443359), despite their distinct TME. The median survival of the 4 groups had similar trend as OS (HL: 8.682192 years; HH: 4.194521 years; LL: 4.186301 years; LH: 2.857534 years).

**Validation in Multidata Sets and Meta-Analysis**

A total 1045 patients from 5 independent data sets with transcriptome and clinical data were used to validate our classification methods on molecular characteristics and prognostic effects. Higher immune score groups showed more immune mediator in all validation data sets (IFN-γ, IL-1,2,6,10; P < 0.01, Supplementary Fig. S11-15). PD-1 activation was observed in high immune groups (P < 0.01), except GSE68571 (P = 0.754). The abundance of different cell population was comparable in GSE68571 (Supplementary Fig. S15), due to less mapping genes (5545 genes vs. more than 13,000 genes in other data sets).

All data sets were used to validate the 166 TME characteristic genes used in our bi-classification (the high an low immune-stromal ratio groups) of NMF consensus clustering. Four data sets showed significant better OS in the high immune-stromal ratio group (P < 0.05), and the 2 remaining data sets (GSE37745 and GSE68571) shared similar trend (Fig. 4A). Meta-analysis were performed for OS in all 6 clinical cohorts for comprehensive analysis of the different results. Because of low heterogeneity (tau^2 = 0, I^2 = 0.0% , P = 0.4673), fixed effect model was selected for meta-analysis in bi-classification groups. The high immune-stroma ratio group had significant better OS in meta-analysis (HR = 0.63, 95%CI: 0.54-0.73, Fig. 3C). There was no obvious bias in all data sets according to linear regression test for funnel plot asymmetry (P = 0.3014, Fig. 3D).

In multidata sets, univariate Cox regression revealed the positive prognostic significance of immune score and the negative prognostic significance of stromal score (Table 2). Survival analysis showed that HL group in the validation data sets also had better OS than other 3 groups, but the other subtypes were inconsistent in different data sets. (Fig. 4B).

Single-arm meta-analysis was utilized to evaluate the prognosis of the 4 subtypes for 3-year and 5-year survival. The HL group had the highest 3-year survival rate (0.81, 95% CI:0.73-0.88, Fig. 3E-H), followed by the LL group (0.64, 95% CI: 0.55-0.74) and the other 2 groups (0.59, 95% CI: 0.53-0.66 for HH; 0.59, 95% CI: 0.54-0.64 for LL). Five-year survival rate had similar trend with 3-year survival in all groups (0.66, 95% CI: 0.56-0.76 for HL; 0.51, 95% CI: 0.38-0.64 for LL; 0.48, 95% CI: 0.42-0.53 for LH; and 0.44, 95% CI: 0.37-0.52 for HH, Fig. 3I-L).

**PMBT Based on immune and stromal activation was Used to Predict Prognosis and Immunetherapy Response**

Because of the opposite predictive effect of the immune and stromal scores on prognosis, we attempted to develop a prognostic evaluation equation that integrated these 2 scores based on Cox model:

*Score of Prognostic Model Based on TME (PMBT) = -1.0635 × immune score + 6.4067 × stromal score*

The related code of PMBT was encapsulated as a R package (GitHub website: https://github.com/ZengZihang/PMBT). The manual of PMBT package was available at NAR online. The PMBT score was significantly related to OS (Table 2). The median PMBT of each data sets was used as the threshold in this study. Clearly, PMBT showed significant prognostic value for OS in 5 data sets except GSE37745 (P = 0.334, Supplementary Fig. 16A). Comparing with existing prognostic methods for LUAD, PMBT scores of 3,5-year survival were related to wider areas under the curve than other TME-based indexes, including T cells, B cells, fibroblasts, PD1 signaling, IFN-γ and previous immune and stroma enrichment scores (Supplementary Fig. 16B).

Our novel PMBT was used to predict anti-CTLA-4 and anti-PD-1 response in melanoma immunotherapy data set (41). The immune score of overall immune response (neglect of therapeutic drugs) was positively correlated to the immunotherapy effects ($P < 0.05$, Supplementary Table 15). On the other hand, PMBT score was negatively correlated to the immunotherapy effects ($P < 0.05$, Supplementary Table 15). While none of the 3 scores was correlated with anti-CTLA-4 treatment ($P > 0.05$, Supplementary Table 15), all the 3 scores of anti-PD-1 therapy showed a better predictive effect in correlation analysis (all, $P < 0.05$) and ROC than those of overall treatment. The PMBT and immune scores had better prediction values than the stromal score (Supplementary Table 15).

**Intercellular Communication Networks in TME**

In the TME, a wide range of signal transmission between cell types occurs through the soluble proteins and direct communication. Intercellular communication networks were constructed using the prior information about cell ligands and receptors, as well as TCGA data.

For all LUAD patients, the subnetwork containing CD8+ T cells, B cells, NK cells and fibroblasts was identified by Molecular Complex Detection (MOCDE) method (Fig. 5A). In this subnetwork, the interactions of ligands and receptors in immune cells and fibroblasts were mediated by collagens (COL1A1 and COL1A2), integrin subunits (ITGAL, ITGA1, ITGA2, ITGA4, ITGA5 and ITGA6), C-X-C motif chemokine receptors (CXCR1, CXCR2 and CXCR3) and TNF superfamily members (TNF, TNFSF13 and TNFRSF1A).

Different subnetworks were identified in different subgroups. Both stroma-centered subnetworks (fibroblasts, endothelial cells and laminin, Fig. 5B) and immunity-centered (CD8+ T cells, B cells, CD molecule, interleukin receptors and TNF superfamily members, Fig. 5C) were identified in the HH group. No stroma-centered and 2 immunity-centered subnetworks were identified in the HL group. One was cellular immune subnetwork consisting of CD4+ T cells, CD8+ T cells, NK cells, dendritic monocyte, CD molecules, C-C motif chemokine ligands and HLA-A (Fig. 5D), and the other was humoral immune subnetwork containing B cells, neutrophils, macrophage monocyte, Class I major histocompatibility complex (HLA-B, HLA-C and HLA-G), leukocyte immunoglobulin-like receptors and C-C motif chemokine receptors (Fig. 5E). These results suggested that C-C motif chemokine mediated the communication of cellular and humoral immune in high immunoreactive subtypes absent of stromal activation.

Contrarily, in the LH group, only fibroblast-centered subnetwork containing ITGB4, LAMB1, LAMB3 and LAMC1 was identified in the LH group (Fig. 5F). No subnetwork with more than 5 nodes was identified in the LL group, suggesting its desert-like molecular communication. Our results indicated that the 4 TME groups classified by PMBT had different molecular and cellular characteristics, as well as excellent predictive effects on prognosis and immunotherapy response in LUAD patients.

**Discussion**

TME played an important role in tumorigenesis and development. However, the landscape of TME, especially immune and stromal cell infiltration on LUAD remained to be fully elucidated. Our study suggested more comprehensive understanding by considering both immune and stromal activity simultaneously. Virtual dissection of mixed tumor tissues was realized by signal decomposition algorithm NMF. Through the NMF, as well as other bioinformatics and machine learning methods, we successfully identified the genetic characteristics of immunity and stroma in LUAD. Based on the 167 TME related genes, we classified LUAD into 4 subtypes with different molecular, cellular and prognostic characteristics.

Although the immune checkpoint inhibitor (anti-PD-1/PD-L1) treatment benefited NSCLC patients (42-44), only about ~16% patients had long-term survival under immunotherapy (6, 45). Screening of potentially sensitive population for immunotherapy helped to decrease medical expenses and improve quality of life. By calculating our immune , stromal and PMBT scores, we found that there was a significantly positive correlation between immune scores and immune response rates in anti-PD-1 therapy ($P = 0.022$), but negative correlation with the stromal scores ($P = 0.039$) and PMBT scores ($P = 0.021$). In conclusion, our PMBT was beneficial to predict the response and prognosis of immunotherapy.

TMB have complex effects on tumorigenesis (46). Driver mutations, such as tumor suppressor gene TP53, may increase genomic instability and increase cell proliferation, which may link to unfavorable prognosis (47, 48). On the other hand, passenger mutations may activate the immune responses through the production of neoantigens and thus contribute to prognosis and immunotherapy response. Recent clinical studies revealed significantly different, even opposite, prognostic effect of TMB in NSCLC patients without immunotherapy. In

LACE-Bio-II (LB2) study (49), high TMB group (≥ 8 m/Mb) had better disease free survival (DFS), OS and lung cancer specific survival (LCSS) in 908 NSCLC patients after complete resection with targeted sequencing of 1538 genes, while the low TMB group (< 4 m/Mb) had worse prognosis (DFS: P = 0.007; OS: P = 0.016; LCSS: P = 0.001). However, another clinical study indicated that higher TMB (≥ 62 m/Mb) correlated with worse OS in 90 NSCLC patients only underwent surgery (HR = 6.633, P = 0.0003) (50). TMB was also found to be a poor prognostic factor in multivariate Cox regression (HR = 12.31, P = 0.019) in all stages, especially in stage I NSCLC patients (OS: HR = 7.582, P = 0.0018; DFS: HR = 6.07, P = 0.0072). TMB may not be a very robust prognostic marker due to lack of elaborate consideration of the biological effects of individual mutation, driver and passenger mutations, as well as the interference of TME RNAs when sequencing. In our study, the HH and LH groups had significantly higher TMB (Fig. 2G), suggesting TMB was directly or indirectly related with stromal activation, in addition to the scores of immune activation. Patients with low stromal activation had less TMB (Fig. 2H). The prediction values of TMB and its relationship with TME remain to be studied. Single-cell whole exome sequencing may provide new insights due to higher purity of the tumor samples.

In intercellular communication network, fibroblast-centered networks were identified in the HH and LH groups, suggesting that fibroblasts mediated stromal activation by specific laminin (LAMA2, LAMB1, LAMB3, LAMC1) and collagens (COL1A2, COL18A1, Fig. 5B & F). In addition to cellular immune related network, humoral immune related cells and molecules (Fig. 5D) were also identified in the HH and HL groups, suggesting that the humoral immune network also had value of cell communication in TME.

Overall, we identified both the immune- and stromal-related gene signatures in LUAD and classified the tumor tissues into 4 groups according to the TME rather than the tumor itself. PMBT was constructed by integrating both immune and stromal scores. PMBT showed excellent value to predict the prognosis and immune response. We expected PMBT score to be validated in more data sets to determine the thresholds. In this study, only lung cancer data were used for analysis. Our results need to be further validated and carefully used in pan-cancer.

## Availability

MCPcounter is an open source R package (https://zenodo.org/record/61372).

CIBERSORT is a free tool developed by Newman et al (http://cibersort.stanford.edu).

TIMER is an open source software in website (https://cistrome.shinyapps.io/timer/)

MALSAR is an open source matlab tool (https://github.com/jiayuzhou/MALSAR).

Cytoscape is an open source software platform for visualizing complex networks and integrating these with any type of attribute data (https://cytoscape.org).

RevMan 5 is the software used for preparing and maintaining Cochrane Reviews (https://community.cochrane.org/help/tools-and-software/revman-5/revman-5-download).

WebGestalt is an open tool (http://www.webgestalt.org/option.php).

PMBT is an R package used for data preprocessing, standardization and calculation of immune,stromal, PMBT scores (https://github.com/ZengZihang/PMBT).

Other R packages can be loaded in CRAN (comprehensive r archive network, https://cran.r-project.org) or Biocounductor (http://bioconductor.org).

## Supplementary data

Supplementary Data are available at *Cancer Research* online.

## Acknowledgments

Thank Yuan Luo and Shijing Ma for their contributions to the writing assistance in this paper.

Author contributions: CH-X and YG leaded and supervised the project. ZH-Z designed the analysis thought, and NN-Z and XP-J provided some points of view. JL-L and JR-C carried out data collection and standardization. YP-G, LX-X and XY-L collected various analysis tools. ZH-Z, JL-L, YK-G, LZ-H and JB-R contributed to statistical analysis. Codes and PMBT package were provided by ZH-Z. ZH-Z, JL-L, YG and CH-X wrote the manuscript.

Impact of Tumor Mutation Burden in Patients With Completely Resected Non-Small Cell Lung Cancer: Brief Report. Journal of thoracic oncology : official publication of the International Association for the Study of Lung Cancer. 2018;13:1217-21.

# Tables

## Table 1

| Variable | High-immune and high stromal-group (n=91) | High-immune and low-stromal group (n=162) | Low-immune and high-stromal group (n=162) | Low immune and low-stromal group (n=91) | P-value |
|---|---|---|---|---|---|
| Gender | | | | | 0.5812 |
| Male | 37 (40.7) | 74 (45.7%) | 80 (49.4%) | 44 (48.4%) | |
| Female | 54 (59.3) | 88 (54.3%) | 82 (50.6%) | 47 (51.6%) | |
| Median age (IQR) | 66.5 (57.7-72.2) | 69.4 (62.7-75.0) | 63.3 (56.2-71.6) | 65.3 (60.5-72.0) | <0.001 |
| Mean smoking years (IQR) | 2.89 (2-4) | 2.84 (2-4) | 2.82 (2-4) | 2.65 (2-4) | 0.489 |
| Pathologic T | | | | | 0.009267 |
| T1-2 | 84 (92.3%) | 147 (91.9%) | 135 (83.3%) | 72 (80.0%) | |
| T3-4 | 7 (7.7%) | 13 (8.1%) | 27 (16.6%) | 18 (20.0%) | |
| Pathologic N | | | | | 0.02856 |
| N0-1 | 76 (84.4%) | 142 (91.6%) | 127 (79.9%) | 74 (82.2%) | |
| N2-3 | 14 (15.6%) | 13 (8.4%) | 32 (20.1%) | 16 (17.8%) | |
| Pathologic M | | | | | 0.4085 |
| M0 | 59 (93.7) | 109 (94.8%) | 110 (94.0%) | 61 (88.4%) | |
| M1 | 4 (6.3%) | 6 (5.2%) | 7 (6.0%) | 8 (11.6%) | |
| Pathologic stage | | | | | 0.008491 |
| Stage I-II | 71 (78.9%) | 137 (86.7%) | 119 (74.4%) | 63 (70.0%) | |
| Stage III-IV | 19 (21.1%) | 21 (13.3%) | 41 (25.6%) | 27 (30.0%) | |
| Tumor status | | | | | 0.004681 |
| Tumor free | 52 (69.3%) | 112 (83.0%) | 92 (73.0%) | 45 (60.8%) | |
| With tumor | 23 (30.7%) | 23 (17.0%) | 34 (27.0%) | 29 (39.2%) | |
| Residual tumor | | | | | 0.1793 |
| R0 | 59 (92.2%) | 106 (98.1%) | 116 (95.9%) | 62 (92.5%) | |
| R1-2 | 5 (7.8%) | 2 (1.9%) | 5 (4.1%) | 5 (7.5%) | |

**Table 2**

| Datasets | Immune score | | Stroma score | | PRTM score | |
|---|---|---|---|---|---|---|
| | coef | P | coef | P | coef | P |
| TCGA | -1.7008 | 0.0038 | 7.245 | 3.29E-05 | 1 | 7.28E-06 |
| GSE11969 | -3.005 | 0.0178 | 2.9082 | 0.00142 | 0.4448 | 0.00101 |
| GSE68465 | -0.6216 | 0.23 | 3.896 | 0.0165 | 0.4804 | 0.0167 |
| GSE68571 | -3.07133 | 0.0295 | 1.39E+01 | 0.00595 | 2.0835 | 0.00185 |
| GSE37745 | -1.2924 | 0.0605 | 0.6 | 0.753 | 0.2422 | 0.334 |
| GSE50081 | -0.9187 | 0.39 | 10.641 | 0.000462 | 1.2122 | 0.00138 |

# Figures

## Figure 1

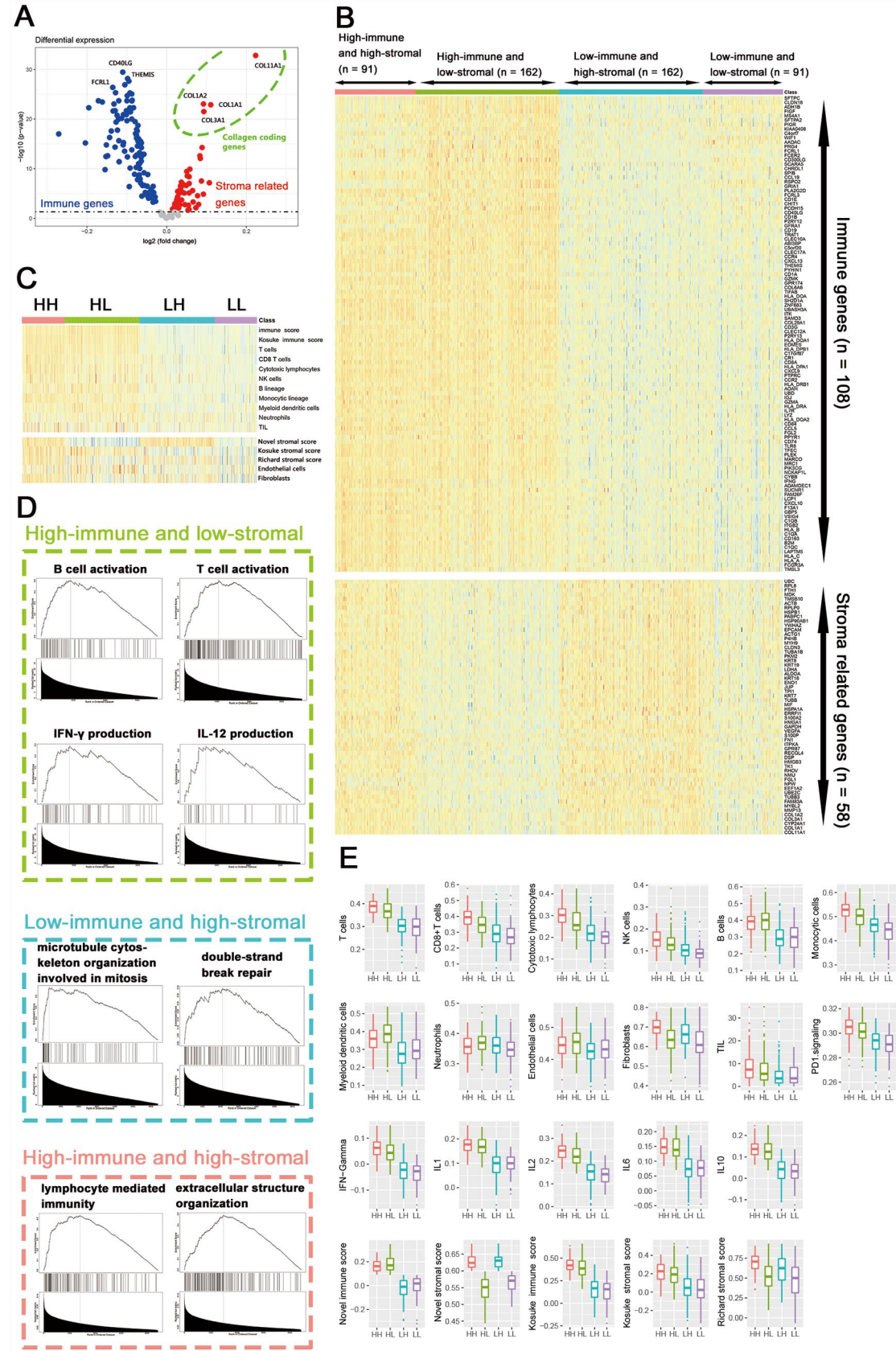

# Figure 2

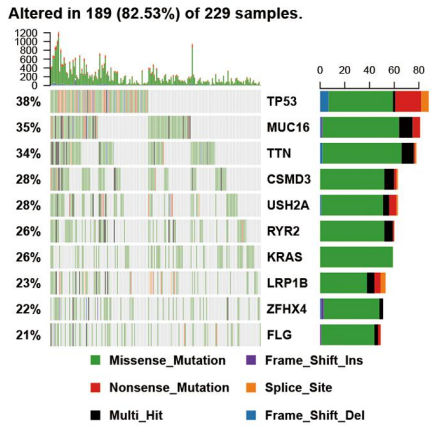
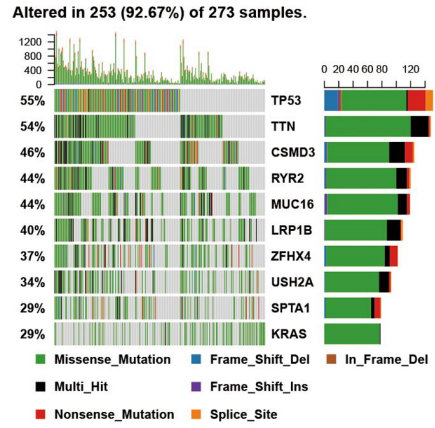
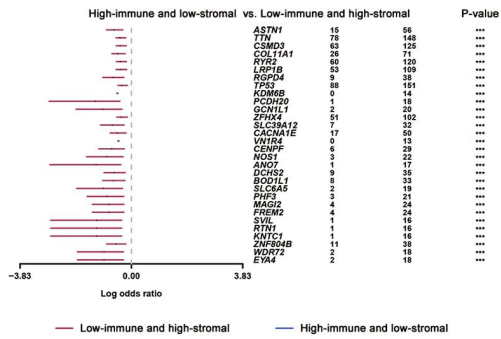
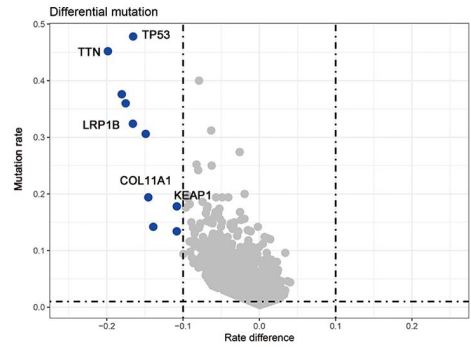
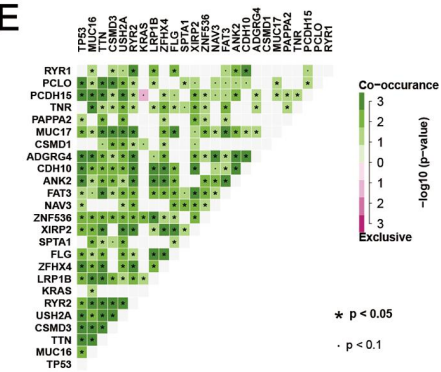
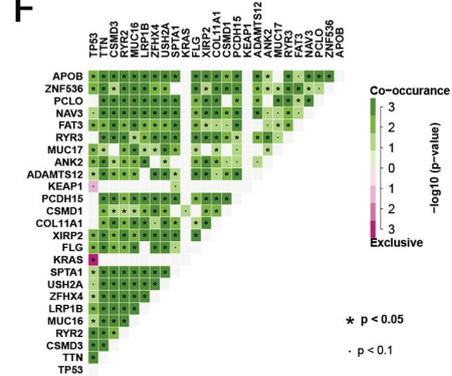
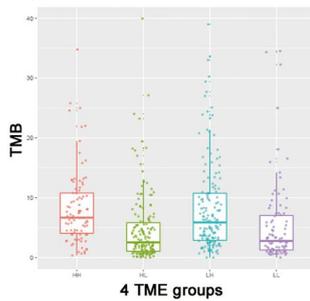
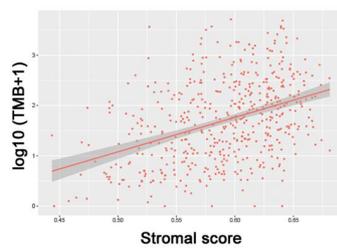
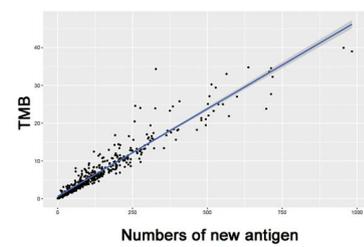

Figure 3

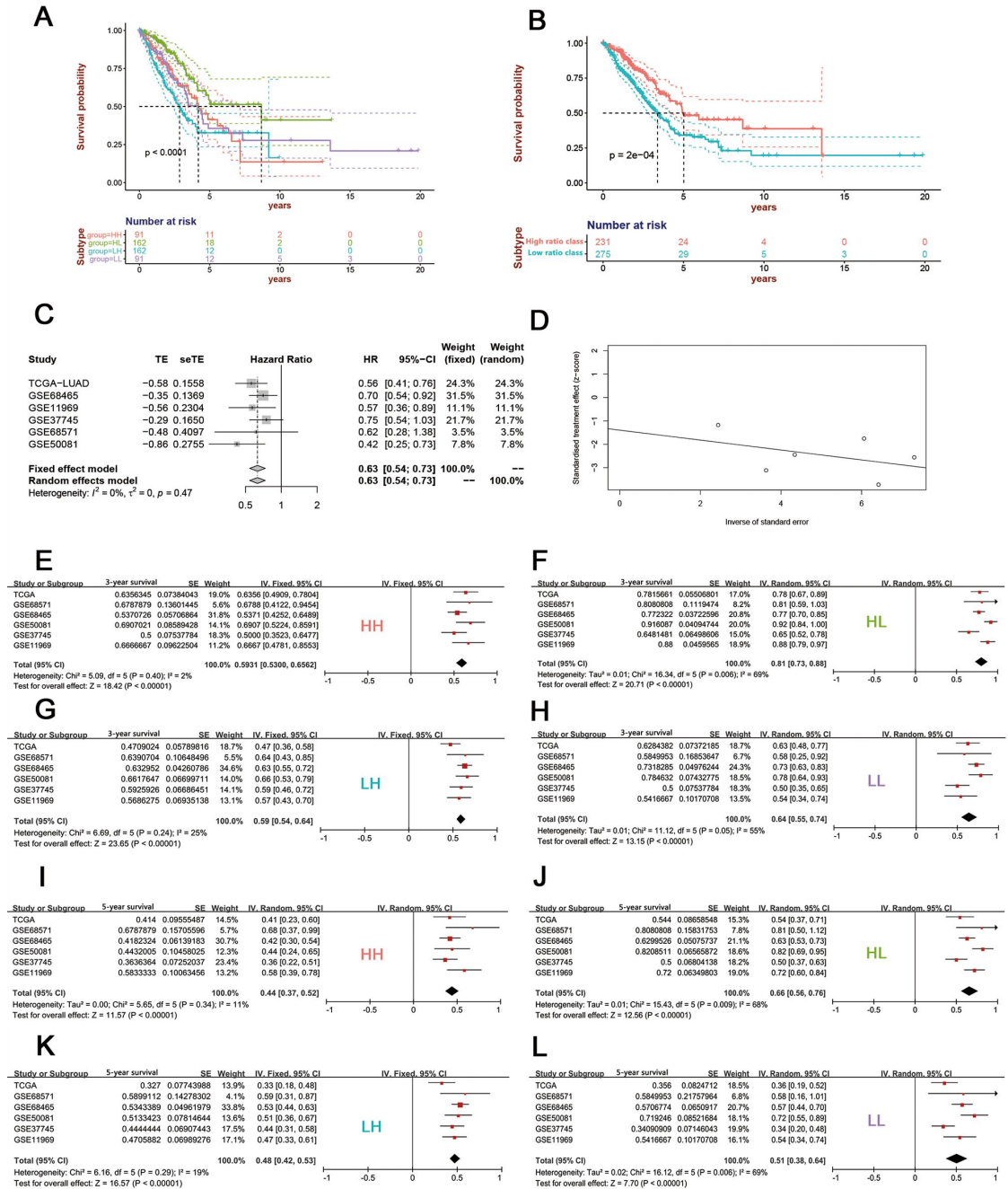

Figure 4

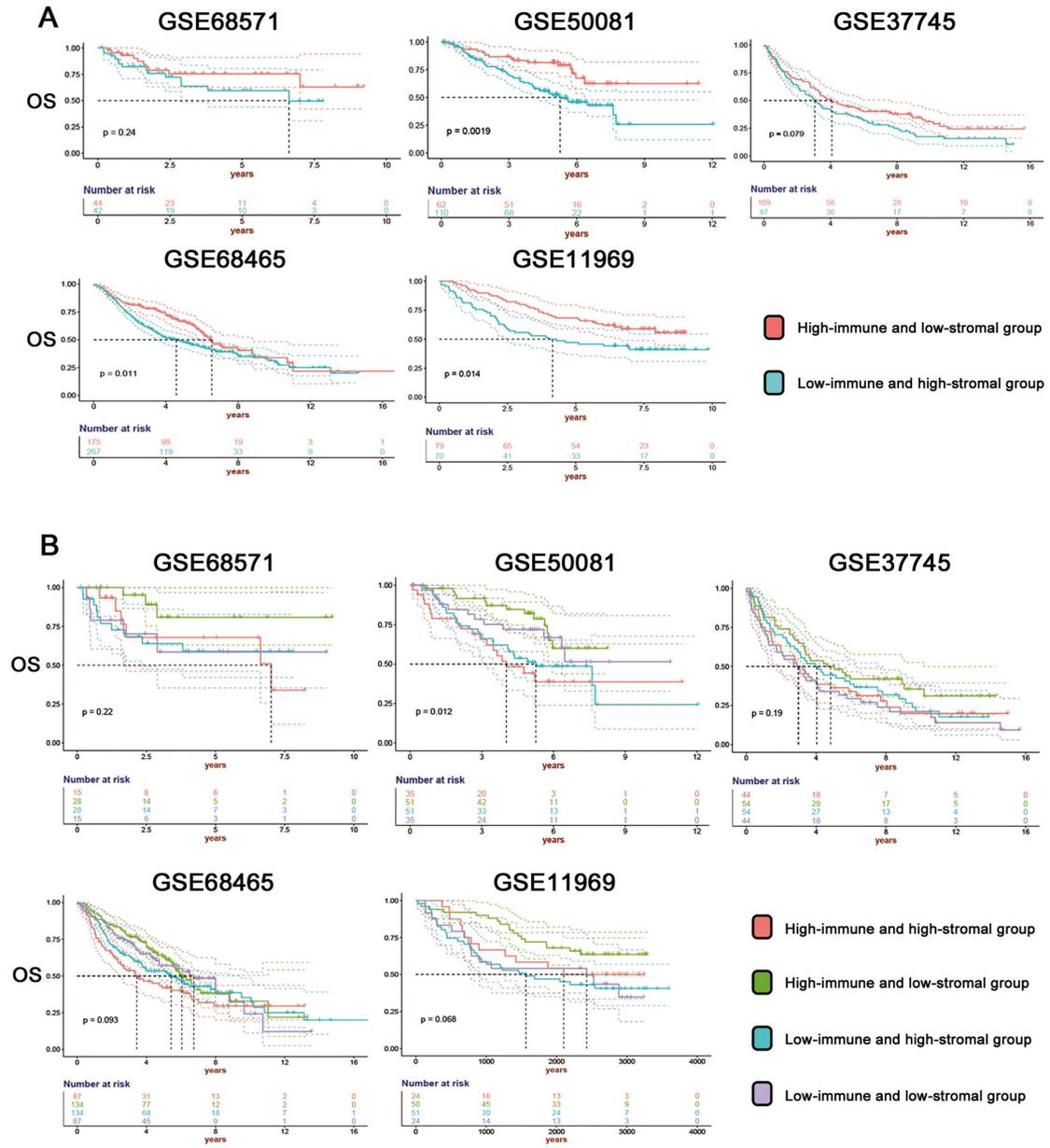

**Figure 5**

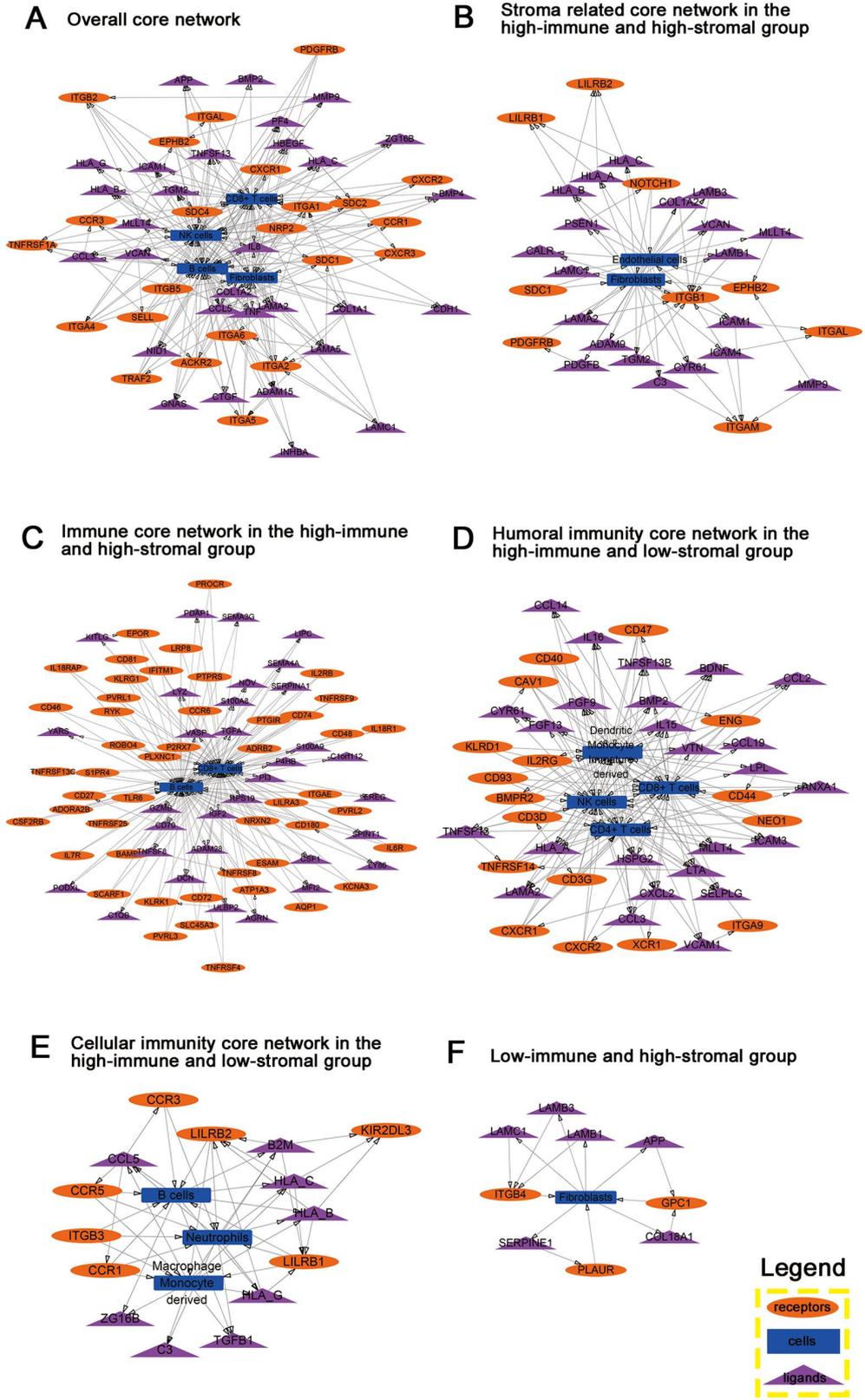

**Table and figure legends**

Table 1. Clinical information table for four subtypes.

Table 2. Univariate Cox regression of immune, stroma and PMBT scores.

Figure 1. Cellular and molecular characteristics in the 4 TME groups based on immune and stroma related gene features. (A) Volcano map of optimized immune and stroma related genes; (B) Heatmap of immune and stroma related genes in the 4 TME groups; (C) Heatmap for cellular characteristics in the 4TME groups; (D) GSEA of the 4 TME groups; (E) Box diagram of cellular and molecular characteristics in the 4 TME groups. The HH meant high-immune and high-stromal group. The HL meant high-immune and low-stromal group. The LH meant low-immune and high-stromal group. The LL meant low-immune and low-stromal group.

Figure 2. The low-immune and high-stromal group had higher driving gene mutation rate and TMB than the high-immune and low-stromal group. (A) Mutation landscape of the high-immune and low-stromal group; (B) Mutation landscape of the low-immune and high-stromal group; (C) forest map of mutation between the high-immune and low-stromal group and the low-immune and high-stromal group; (D) Volcano map of mutation rate between the high-immune and low-stromal group and the low-immune and high-stroma group; (E) Co-occurance and exclusive mutation in the high-immune and low-stromal group; (F) Co-occurance and exclusive mutation in the low-immune and high-stromal group; (G) Box diagram of TMB in the 4 subtypes; (H) Scatter plot between TMB and stromal scores; (I) Scatter plot between TMB and numbers of new antigens. The HH meant high-immune and high-stromal group. The HL meant high-immune and low-stromal group. The LH meant low-immune and high-stromal group. The LL meant low-immune and low-stromal group.

Figure 3. The high-immune and low-stromal group had better survival than other groups. (A) Survival curves of the 4 TME subgroups by ssGSEA; (B) Survival curves of the high and low immune-stroma ratio groups by NMF consensus clustering; (C) Meta-analysis of OS in training and validation data sets; (D) Funnel plot of OS meta-analysis; (E-H) Meta-analysis of 3-year survival rate in the HH, HL, LH and LL groups; (I-L) Meta-analysis of 5-year survival rate in the HH, HL ,LH and LL groups. The HH meant high-immune and high-stromal group. The HL meant high-immune and low-stromal group. The LH meant low-immune and high-stromal group. The LL meant low-immune and low-stromal group.

Figure 4. OS of the 5 validation data sets. (A) Survival curves of the high and low immune-stromal ratio groups; (B) Survival curves of the 4 groups.

Figure 5: (A) Core subnetwork of all patients; (B,C) Core subnetworks of high-immune and high-stromal group; (D,E) Core subnetworks of high-immune and low-stromal group; (F) Core subnetwork of the low-immune and high-stromal group